\documentclass{llncs}

\usepackage{graphicx}
\usepackage{sidecap}
\usepackage{setspace}
\usepackage{amsmath}

\sidecaptionvpos{figure}{c}
\sidecaptionvpos{table}{c}

\usepackage{url}
\usepackage{nicefrac}
\usepackage{hyperref}

\usepackage{enumitem}
\setlist{nolistsep}

\usepackage{fancyhdr}
\fancypagestyle{firststyle}
{
    \fancyfoot[CE,CO]{\scriptsize{
        Preprint submitted to ARC 2015.\\
        \href{http://link.springer.com/chapter/10.1007\%2F978-3-319-16214-0_29}{The final publication is available at link.springer.com.}
    }}
}

\graphicspath{{figuras/}}
\DeclareGraphicsExtensions{.pdf,.jpeg,.png}

\begin{document}

\title{Modular Acquisition and Stimulation System for\\Timestamp-driven Neuroscience Experiments}

\author{Paulo Matias \and Rafael T. Guariento \and \\
Lirio O. B. de Almeida \and Jan F. W. Slaets}
\institute{S\~ao Carlos Institute of Physics\\
University of S\~ao Paulo\\
S\~ao Carlos, SP, Brazil\\
\email{\{matias,guariento,lirio,jan\}@ifsc.usp.br}
}

\maketitle
\thispagestyle{firststyle}
\begin{abstract}
Dedicated systems are fundamental for neuroscience experimental protocols that require timing determinism and synchronous stimuli generation. We developed a data acquisition and stimuli generator system for neuroscience research, optimized for recording timestamps from up to 6~spiking neurons and entirely specified in a high-level Hardware Description Language (HDL). Despite the logic complexity penalty of synthesizing from such a language, it was possible to implement our design in a low-cost small reconfigurable device. Under a modular framework, we explored two different memory arbitration schemes for our system, evaluating both their logic element usage and resilience to input activity bursts. One of them was designed with a decoupled and latency insensitive approach, allowing for easier code reuse, while the other adopted a centralized scheme, constructed specifically for our application. The usage of a high-level HDL allowed straightforward and stepwise code modifications to transform one architecture into the other. The achieved modularity is very useful for rapidly prototyping novel electronic instrumentation systems tailored to scientific research.

\end{abstract}

\section{Introduction}
Neurons usually behave by emitting stereotyped pulses of electric depolarization through their membranes, creating temporally localized spikes. It is a common belief that spiking neurons follow an all-or-none principle, similar to the processing of digital signals, by encoding information only through spike timing \cite{Dayan:2005:TNC:1205781}. Although each individual cell always produces the same waveform, the most widespread experimental approach employs Analog to Digital Converters (ADCs) integrated on commercial acquisition systems to capture complete waveforms. This procedure is required when the researcher desires to analyze a large population of neurons recording only from a few electrodes, applying then a neuron classification technique known as spike sorting to discriminate individual waveforms \cite{reviewSpkSort}. However, because of the lack of readily available specialized acquisition hardware, many works adopt the same recording technique even though they only need to identify the occurrence of spikes from one neuron per electrode \cite{Brochini24082011,10.1371/journal.pcbi.1000860,Bolzon11112009,spiketimingauditory}. The resulting data files are large and spikes need to be detected by software, demanding a considerable amount of time. 

In this paper we present the design of a low-cost alternative hardware solution based on a dedicated Complex Programmable Logic Device (CPLD). We have chosen CPLDs instead of Field Programmable Gate Arrays (FPGAs) to demonstrate the flexibility of our approach, as CPLDs are usually limited to a small number of logic gates, and lack common FPGA features such as Block RAMs and Phase Locked Loops (PLLs). We implemented the logic circuits on the CPLD adopting a modular design, which aims to facilitate future refinement and customization for specific applications. The complete source code implemented in the Bluespec SystemVerilog (BSV) \cite{1459818} language is available at \cite{Matias:11034}.

BSV designs targeted at small reconfigurable devices, such as ours, are rare in literature, since many works show that BSV usually produces a higher logic element (LE) count than Reg\-is\-ter-Trans\-fer Level (RTL) languages \cite{4438138,overviewESL,Malik:2012:ERA:2460216.2460228}. However, some research \cite{Arvind2004HSE} argues that microarchitectural choices have greater impact on the LE usage than the specification's abstraction level, although there is a lack of studies in glue logic sized architectures with significant modularity and complexity. This paper showcases such a system, and also explores the impact of latency insensitive module decoupling \cite{5762736},
by comparing two distinct implementations of a resource arbitration scheme. Similar work evaluating synthesis results exists \cite{Murray:2014:QCB:2554688.2554786}, but we also test the consequences on system resilience to extreme conditions, many times above our application requirements.

The acquisition input is provided to our digital logic by an analog front-end system which generates an asynchronous TTL-compatible signal pulse at the occurrence of each valid spike. Our entire circuit was designed to be compatible and easily inserted into a previous experimental setup \cite{deAlmeida20111762} devised for studying neural codification in \textit{Chrysomya megacephala}'s visual system, but it is sufficiently generic to be suitable for a wide range of neuroscience experiments.

\subsubsection*{Main contributions of this work:}
\begin{itemize}
    \item{Develops a portable, low-cost and precise data acquisition system for neuroscience and neuroethology experiments.}
    \item{Applies the seldom used concept of recording digital events (instead of ADC-converted data) to increase the precision of neural spike timing.}
    \item{Employs the BSV language in a small and resource constrained system.}
    \item{Showcases architecture refactoring from a decoupled to a centralized scheme.}
\end{itemize}

\subsubsection*{Paper organization:} The next section describes the basic specifications of our design and its overall architecture. Section \ref{BSVmod} discusses the system implementation, focusing on points common both to dynamic and static arbiter versions. Sections \ref{Dynamicarb} and \ref{Staticarb} delve into specific aspects of each one of the implementations. Section \ref{Results} presents synthesis, experimental and simulation results. Finally, we conclude in Section \ref{Conclusion}.

\section{Overall system architecture}\label{Overall}

Our system offers 6~TTL-level pulse timestamp acquisition inputs, 4~analog 16-bit resolution outputs for stimuli generation and a Join Test Action Group (JTAG) host computer interface. It is composed by a MAX II Micro Kit (EPM\-2210\-F324\-C3 CPLD), a 74HC4050 buffer for input overvoltage protection, a MAX5134 Digital-to-Analog Converter (DAC) and a IDT71256 20ns 32K$\times$8-bit SRAM. We have divided the project in following functional subunits:

\textbf{Synchronizer}: Receives asynchronous input pulses and registers 32-bit timestamps from a hardware counter, each one paired to a flag indicating which input channels fired since last counted. In most neural systems, 1~$\mu$s is believed to be enough resolution for studying fine details of information coding \cite{10.1371/journal.pcbi.1000025}.

\textbf{FIFO SRAM}: Provides an interface for using the external SRAM memory as a pair of First-In First-Out (FIFO) queues of 16~KiB each. One of them buffers data acquired from inputs, and the other buffers stimuli received from a computer. Our FIFO modules are compatible with the BSV standard library.

\textbf{JTAG interface}: Provides communication with a host computer. We have wrapped Altera JTAG-UART libraries into a ready-to-use BSV module. By using this protocol, the same communication module is portable to any CPLD or FPGA manufactured by the same vendor. As programmable devices are configured via JTAG, the bus is readily available through USB adapters embedded in almost every evaluation board. However, this approach introduces a significant protocol overhead by encapsulating UART emulation inside JTAG-USB, limiting the data rate to about 1~Mbit/s. Also, client software needs to explicitly poll the device, because the interface is not interrupt nor event driven. This results in software determinism becoming a bottleneck depending on hardware buffer size and desired data rate. Nevertheless, these limitations do not impair this particular application.

\section{BSV module architecture}\label{BSVmod}

Bluespec SystemVerilog is a strongly typed high-level hardware description language (HDL) with functional paradigm features. A BSV design is organized in modules and rules. Modules provide interfaces, composed by a set of methods which can be used to access or modify their internal state. State changes (side-effects) are clearly separated from read-only operations by the means of a monad \cite{WadlerMonad} called \texttt{Action}, thus any expression which modifies state has an action type. Modules can be statically elaborated several times, allowing to represent complex circuit structures. Rules are entities which describe the connections between modules and ultimately define hardware dynamics. They are formed by a set of actions and a boolean predicate, which defines an explicit condition needed to allow execution of the actions (rule firing). During a single clock cycle, a rule is guaranteed to entirely complete its execution or not to fire at all, property known as transaction atomicity. Rule firing can also be affected by implicit conditions, which can be attributed to any BSV expression. The BSV compiler propagates an implicit condition back to the predicate of the rule which actually executes the action or queries the value of the corresponding expression. Implicit conditions are usually attributed to method boundaries, serving as an effective way to specify module contracts. When synthesizing, the compiler defines an execution order for rules, allowing a hardware scheduler to be generated according to the Term Rewriting Systems (TRS) formalism \cite{Shen98designand}.

\begin{figure*}[htbp]
\centering
\includegraphics[width=4.5in]{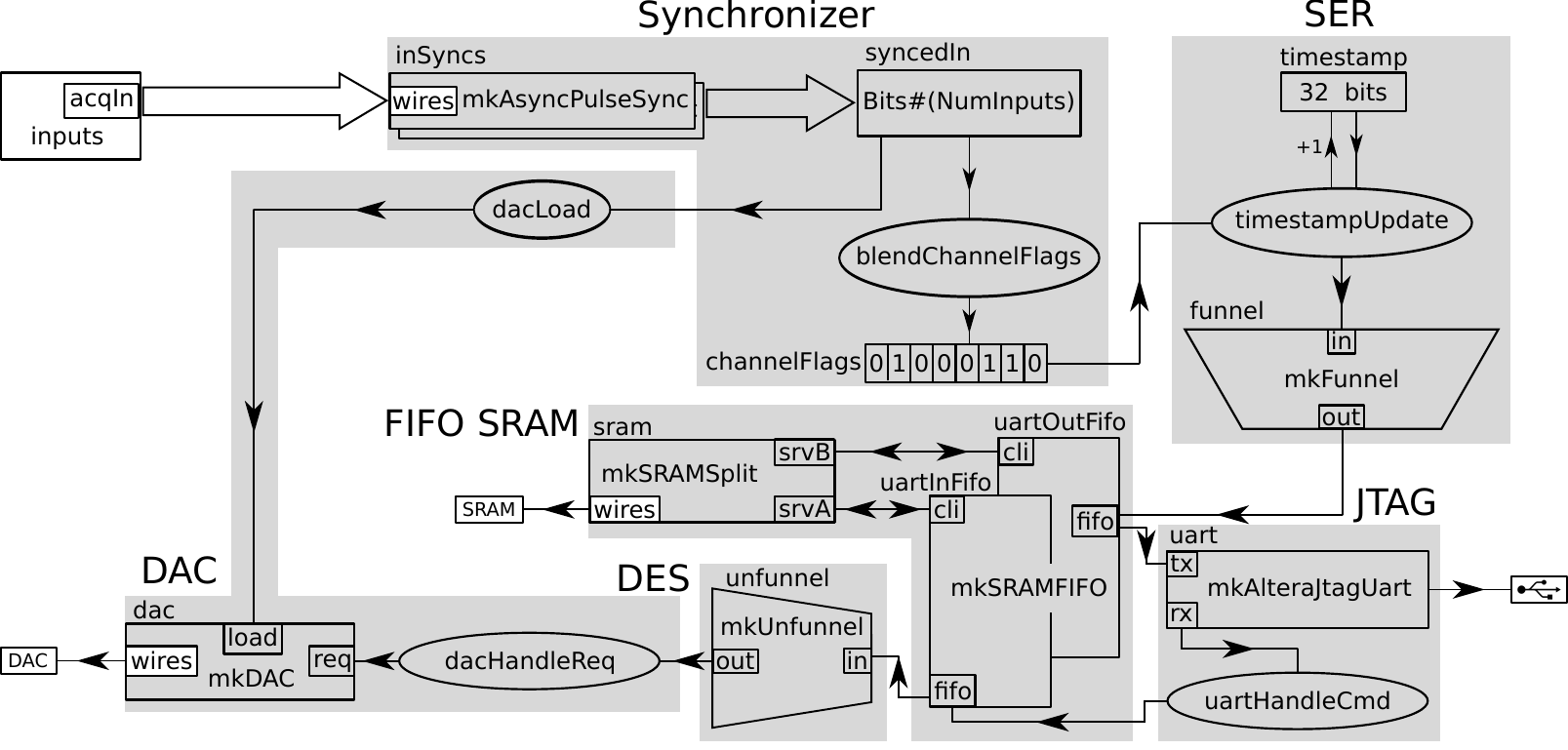}
\caption{Block diagram of the BSV modules and rules. The gray shaded areas represent internal logic, while white structures depict I/O interfaces. Modules are portrayed as quadrilaterals. The names inside them are module names, and those outside are instance names. Small rectangles on their side are interfaces. Ellipses designate rules. Those which only perform connections are omitted and represented directly by arrows.}
\label{BlocosAcqSysDyn}
\end{figure*}

Figure~\ref{BlocosAcqSysDyn} illustrates our main module. The fundamental difference between arbitration approaches resides on \texttt{SRAMFIFO} internals and on how communication with the SRAM occurs. In the dynamic approach this is done by a client-server interface mediated by FIFOs, whereas in the static one an extra central module exists which assigns a specific operation to the \texttt{SRAMFIFO}s on each cycle. 

\textbf{Acquisition data flow}: Asynchronous pulses arriving from acquisition inputs are synchronized to the system clock by the \texttt{AsyncPulseSync} modules, resulting in the \texttt{syncedIn} signal. The \texttt{blendChannelFlags} rule accumulates one bit for each input channel in the \texttt{channelFlags} register, if a pulse occurred on \texttt{syncedIn} since the last collected timestamp. The \texttt{timestamp\-Update} rule atomically increments the timestamp register, sends the \texttt{channelFlags} value and the current timestamp to the funnel, and resets \texttt{channelFlags} to zero. The funnel  emits one byte of its input per cycle to the \texttt{uartOutFifo}. Finally, data coming from \texttt{dataOutFifo} can be read in the host computer after being collected by the JTAG-UART transmitting (\texttt{tx}) interface.

\textbf{Stimulus generation data flow}: Begins at the JTAG-UART receiving (\texttt{rx}) interface. The \texttt{uartHandleCmd} rule identifies if the byte received from the computer represents a start command or a DAC conversion request. A start command sets a boolean register (omitted from the figure) which unblocks the predicate of \texttt{dacLoad}, \texttt{timestampUpdate} and \texttt{blendChannelFlags} rules. A DAC conversion request sends the current byte and the next two bytes to \texttt{uartInFifo}. After coming out of the FIFO, the bytes feed the unfunnel block, merging three bytes together. The \texttt{dacHandleReq} controls the request flow to the DAC module. The \texttt{dacLoad} rule fires when the first input channel receives a pulse, unblocking the \texttt{dacHandleReq} rule and causing a synchronous update on all DAC outputs. This input channel is used to synchronize analog outputs to the desired stimuli clock, e.g. the display controller in a visual stimulation system.

BSV has an implicit condition mechanism which eases the specification of a provably correct system. We only needed to add error handling to four places of our design. The first one is related to \texttt{tx} path FIFO overflow and is put in the \texttt{timestampUpdate} rule, ensuring that the timestamp is always incremented at each update period. The second check is accomplished in the \texttt{dacLoad} rule, and verifies if a FIFO underrun has occurred in the \texttt{rx} path, by certifying that the DAC is ready to receive a new command and that all DAC registers were filled since the last load. The final two are not directly related to system functional correctness, but to good debugging practices. The third one checks if bytes received from JTAG-UART correspond to valid commands. The last one verifies if DAC requests are still valid after leaving \texttt{uartInFifo}, aiming to detect any occurrences of data corruption during communication with the SRAM chip. When any of these error condition occurs, we alert the user by blinking LEDs until the system is reset.

Next, we describe characteristics of the common system sub-modules.

\begin{figure}[hbtp]
\begin{center}
\boxed{\includegraphics[width=2.2in]{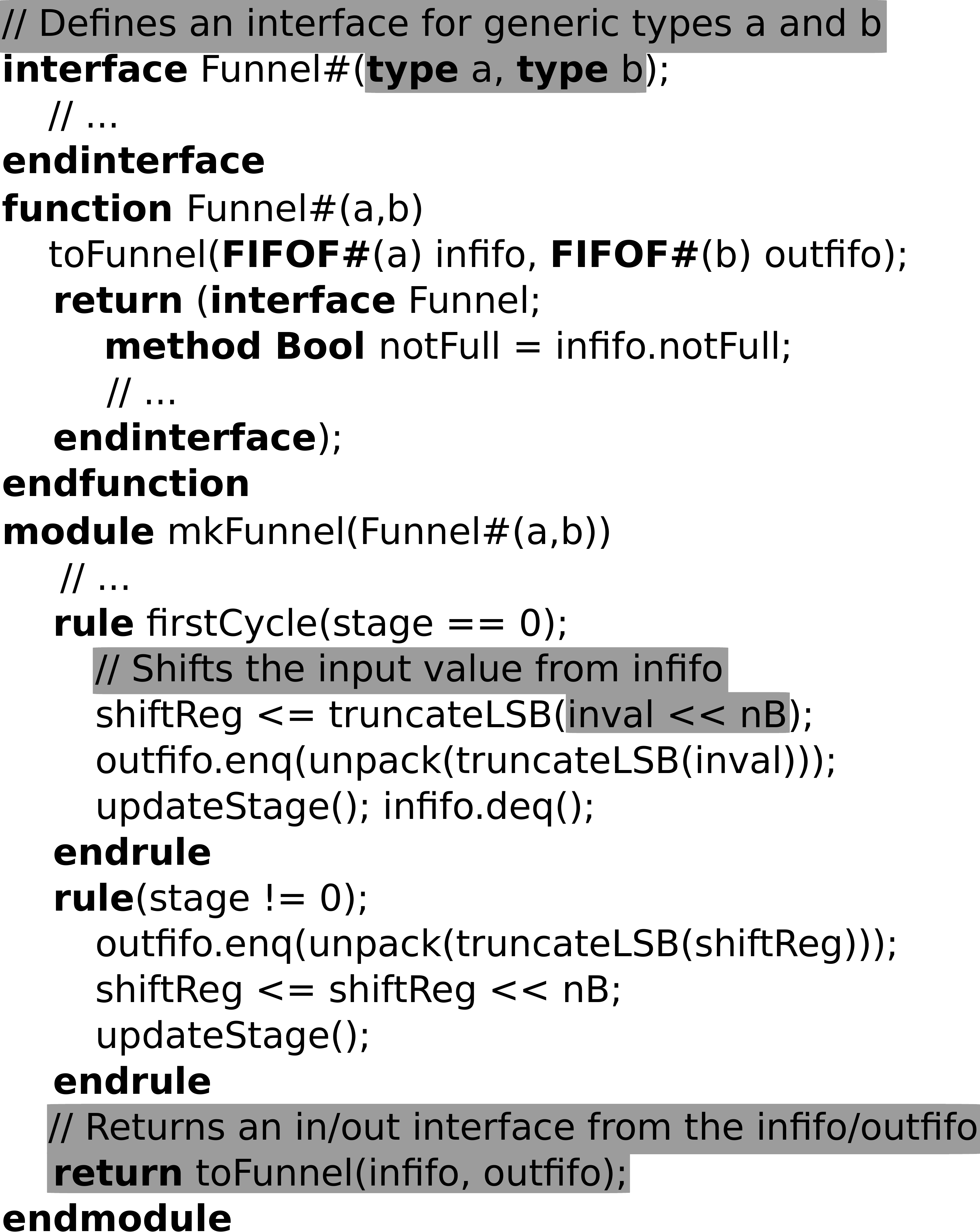}}
\caption{Code excerpt from the serializer shift register implementation, illustrating parametrized data types and abstract manipulation of interfaces by compile-time resolved functions (\texttt{toFunnel}).}
\label{SERDES}
\end{center}
\end{figure}

\textbf{SER/DES}: Serializer and deserializer modules are implemented using shift registers. Our design is generic and type parametrized, making it reusable with any input or output data types. A code excerpt illustrating these concepts is shown in Figure~\ref{SERDES}.

\begin{figure}[htbp]
\begin{center}
\boxed{\includegraphics[width=2.2in]{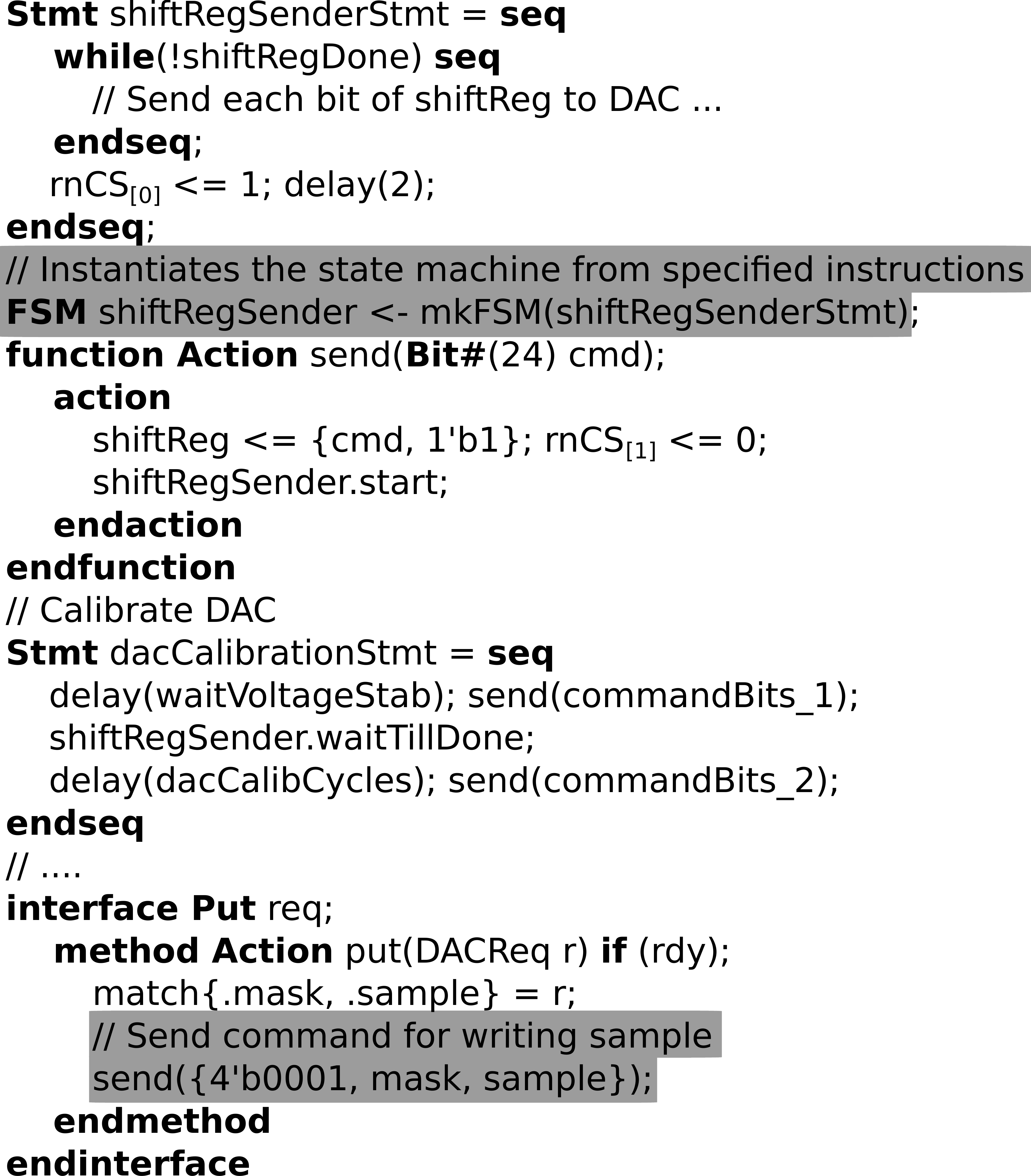}}
\caption{Code illustrating usage of the Finite State Machine (FSM) Domain Specific Language (DSL). FSMs can be specified as a series of statements (\texttt{Stmt}) in a DSL which resembles an imperative software language. The \texttt{mkFSM} module transforms these statements into a hardware implementation. We have implemented a \texttt{send} function which is used both to compose different FSMs together and to start a FSM from an externally accessible method.}
\label{DAC}
\end{center}
\end{figure}

\textbf{DAC}: In order to rapidly prototype the control of a Serial Peripheral Interface (SPI) and DAC linearity calibration procedures, we employed a standard BSV library called \texttt{StmtFSM}, which consists of a Domain Specific Language (DSL) for specifying Finite State Machines (FSMs). The FSMs could be easily composed and exposed in the form of a simple external module interface. DAC register update requests supported by the MAX5134 DAC contain three bytes, one specifying the target channels and two bytes of data. Part of the FSM implementation is illustrated in Figure~\ref{DAC}.

\textbf{SRAMFIFO}: This module is also type parametrized. It exposes a \texttt{fifo} subinterface mimicking BSV standard library's \texttt{mkLFIFO}, and a \texttt{cli} subinterface which can be connected to a \texttt{SRAM} or \texttt{SRAMSplit} server interface. Usage of Ephemeral History Registers (EHRs) \cite{1459853} greatly simplified the design in order to achieve the same scheduling specifications as \texttt{mkLFIFO}. EHRs provide a register-like interface on which same-cycle accesses can be ordered according to the logical execution order of rules or methods. Head and tail position pointers to locations inside the SRAM are held in EHRs. The \texttt{SRAMFIFO} stores one unit of data (in the case of this design, one byte) in a local cache FIFO implemented using flip-flop registers, whose output is connected directly to \texttt{SRAMFIFO}'s one. When the cache FIFO goes empty, we dispatch a read request to the SRAM, aiming to maintain the cache filled most of the time. When new data is enqueued to the \texttt{SRAMFIFO} and no space is available in the cache FIFO, a write request is sent to the SRAM.

\subsection{Dynamic arbiter}\label{Dynamicarb}

\begin{figure*}[htbp]
\centering \includegraphics[width=4.7in]{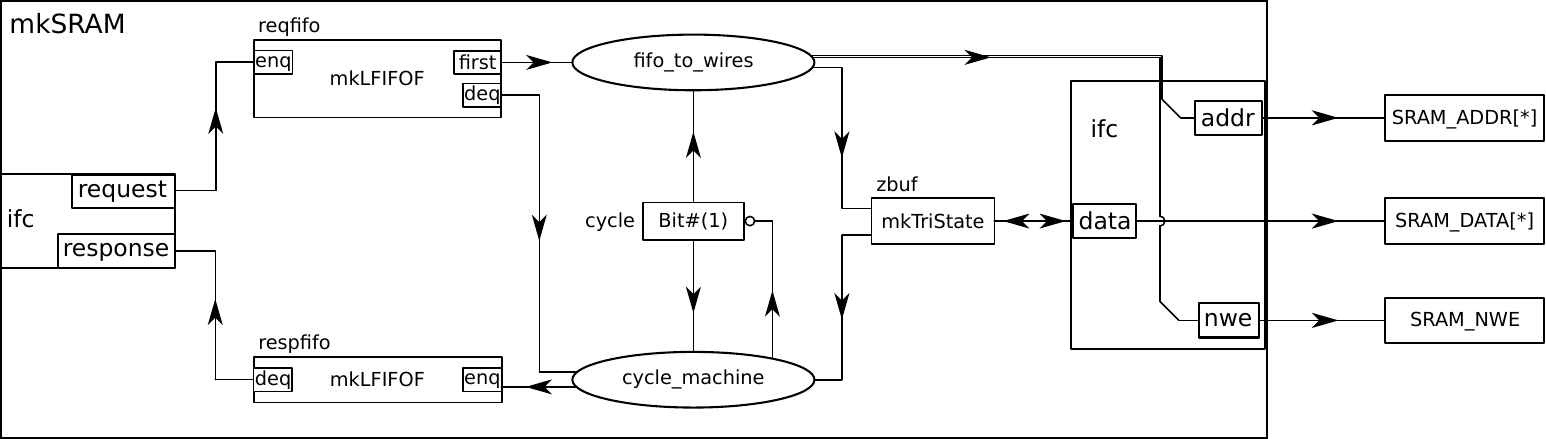}
\caption{In the dynamic arbiter based design, the SRAM controller is decoupled from the \texttt{SRAMSplit} module. Its state over the two cycles of operation is determined by an internal \texttt{cycle} register, handled by the \texttt{cycle\_machine} rule. The \texttt{fifo\_to\_wires} rule controls the tri-state buffer and drives the outputs.}
\label{BlocosSRAM2CLKDyn}
\end{figure*}

\begin{figure}[htbp]
\centering
\includegraphics[width=4.2in]{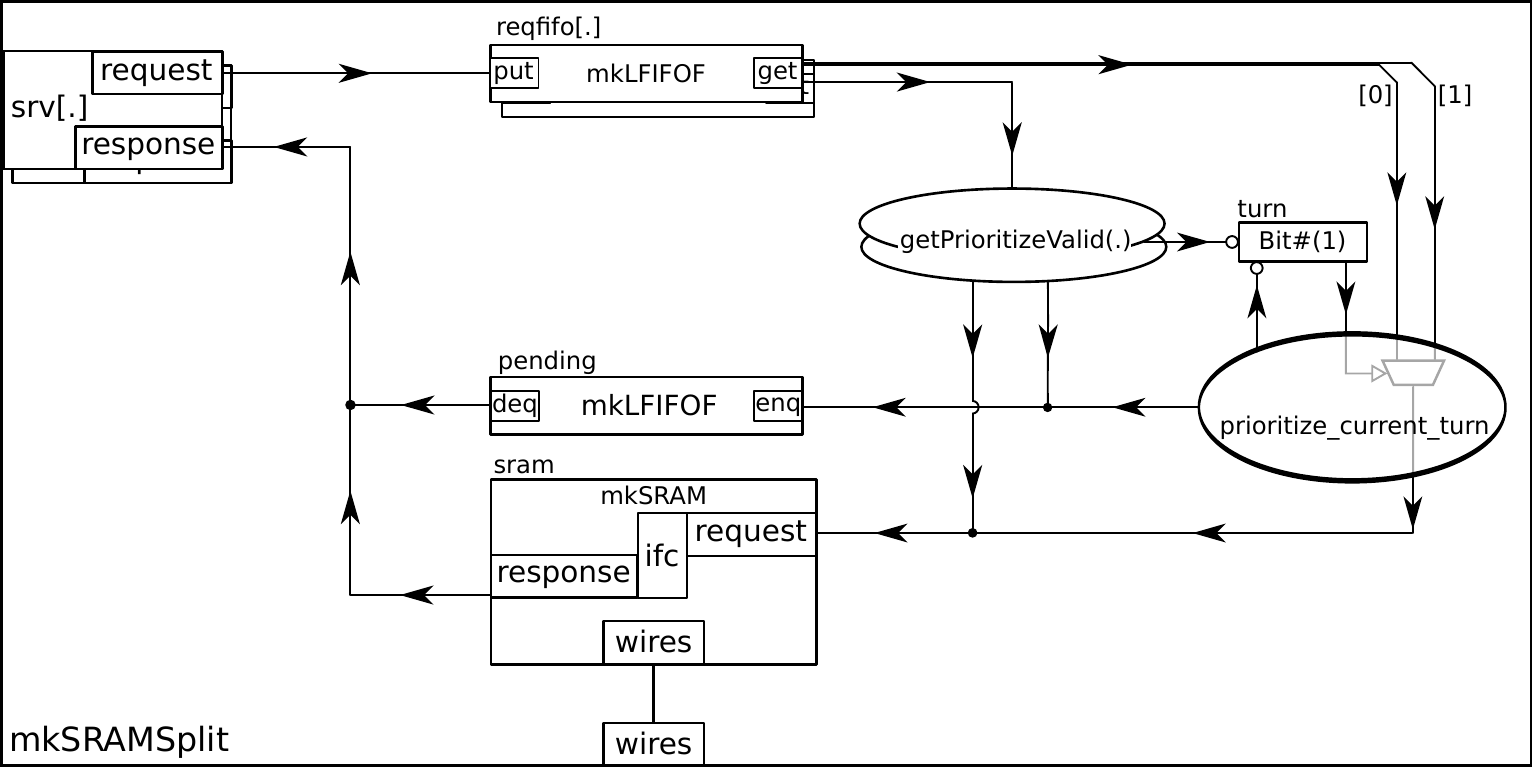}
\caption{\texttt{SRAMSplit} provides two \texttt{SRAM} server interfaces and arbitrates their access to a single SRAM controller. Requests coming to both servers are held into FIFOs until being sent to the controller. A set of three mutually exclusive rules arbitrate the access based on not-empty FIFO flags and on a \texttt{turn} register. A \texttt{pending} FIFO preserves the requester identification, allowing responses to be served back in the right order.}
\label{BlocosSRAMSplitDyn}
\end{figure}

In this design, the SRAM controller (Figure~\ref{BlocosSRAM2CLKDyn}) is decoupled from the \texttt{SRAMSplit} module (Figure~\ref{BlocosSRAMSplitDyn}). The first dispatches requests in the order as they are received in \texttt{reqfifo}, using an internal \texttt{cycle} register to keep track of its state during a single request. The latter arbitrates the access of two other modules to a single SRAM controller. Requests received from both modules are placed into a pair of FIFOs. A set of mutually exclusive rules then controls the priority of each request. Two of them are generated by the \texttt{get\-Pri\-or\-i\-tize\-Val\-id} function, one of which is fired when a single request FIFO is not empty. However when both FIFOs contain data, the \texttt{pri\-or\-i\-tize\_\-cur\-rent\_\-turn} rule is fired, prioritizing the Least Recently Used (LRU) FIFO. The \texttt{turn} register holds the state needed to infer the LRU FIFO.

\begin{figure*}[htbp]
\begin{center}
\boxed{\includegraphics[width=4.6in]{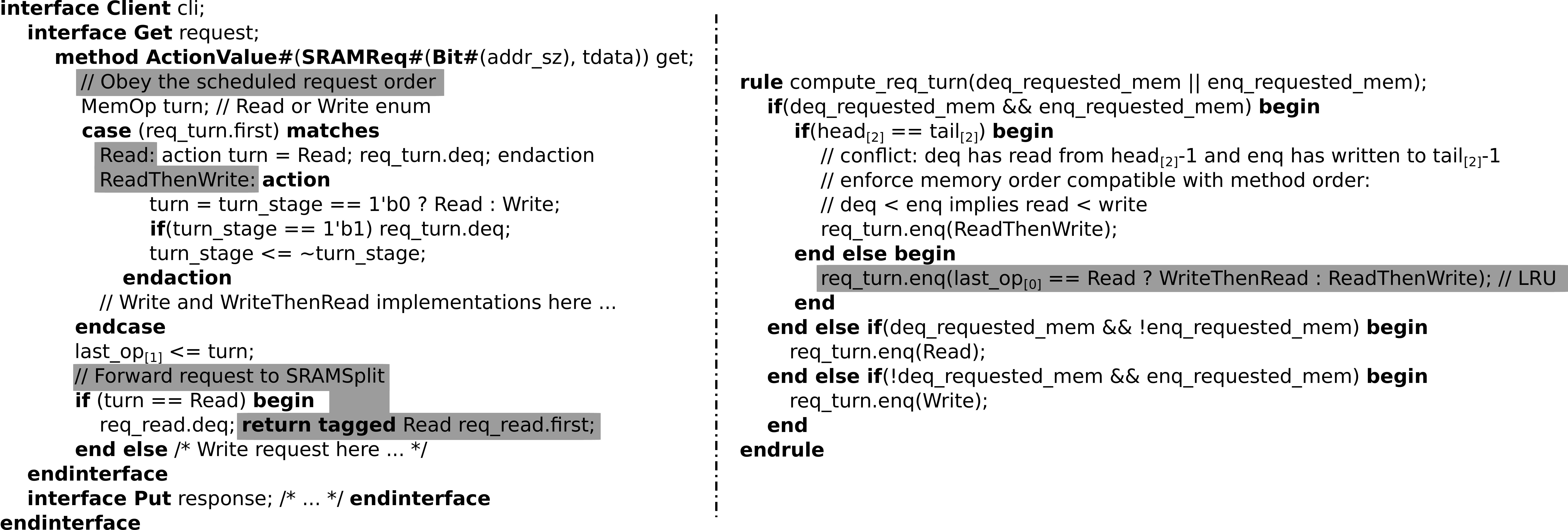}}
\caption{Excerpts from the dynamic arbitrated \texttt{SRAMFIFO} implementation. All of our modules follow a client/server pattern, demanding the treatment of simultaneous requests for full decoupling. The \texttt{compute\_req\_turn} rule (right panel) chooses the turn according to which requests were issued at the current cycle. Whenever possible, a Least Recently Used (LRU) scheme is adopted (highlighted in the code). The \texttt{request.get} method of the client interface (left panel) queries this information from the \texttt{req\_turn} FIFO, updates \texttt{last\_op} and forwards the correct request by returning it. The \texttt{last\_op} is an example of an Ephemeral History Register (EHR) --- references to it must be appended with an index (shown as subscript text in the code) which defines the logical execution order of register read and write operations.}
\label{DynamicArbiter}
\end{center}
\end{figure*}

Arbitration also needs to take place in the \texttt{SRAMFIFO} module, because methods for enqueuing and dequeuing data are designed not to conflict, in order to simplify module reuse. When both methods are called during the same cycle, we check if the queue's head and tail pointers are equal to each other. This means that the dequeue method has requested to read the same address which the enqueue method asked to write. In this case, we enforce requests to be sent to SRAM in the same order as the logical execution order chosen when designing the methods (dequeue before enqueue, and thus read before write). Otherwise, we follow a LRU approach based on the value of a register which holds the type of the last memory operation issued by the \texttt{SRAMFIFO} module (Figure~\ref{DynamicArbiter}).

\subsection{Static arbiter}\label{Staticarb}

Starting from the code of the dynamic version, we incrementally added new conditions to method predicates, testing the system after the changes. As implicit conditions which control the data flow of FIFOs are still present in the logic at this development stage, designer errors tend to prevent rules from firing, stopping data flow and making the system hang instead of producing incorrect results.

\begin{SCfigure}[][htbp]
\centering
\includegraphics[width=2.5in]{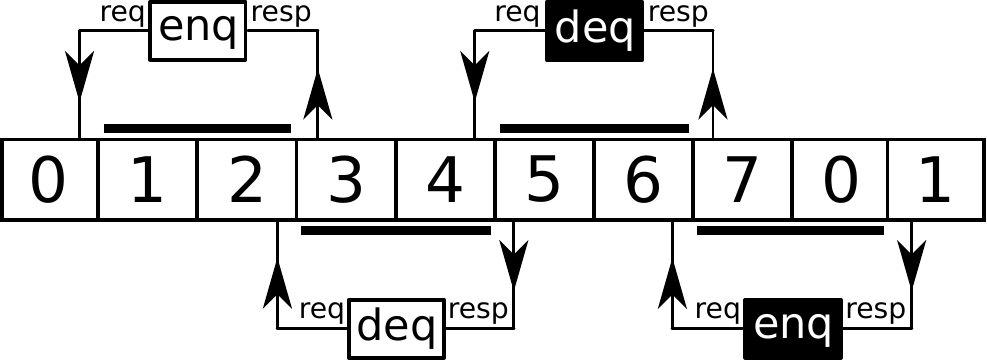}
\caption{Timing diagram of \texttt{SRAMFIFO} transactions governed by the static arbiter. Arrows identify in which cycles the memory requests and responses occur. White rectangles represent actions on \texttt{uartInFifo}, while black rectangles depict operations on \texttt{uartOutFifo}.}
\label{DiagramaTempo}
\end{SCfigure}

\begin{figure*}[htbp]
\begin{center}
\boxed{\includegraphics[width=4.6in]{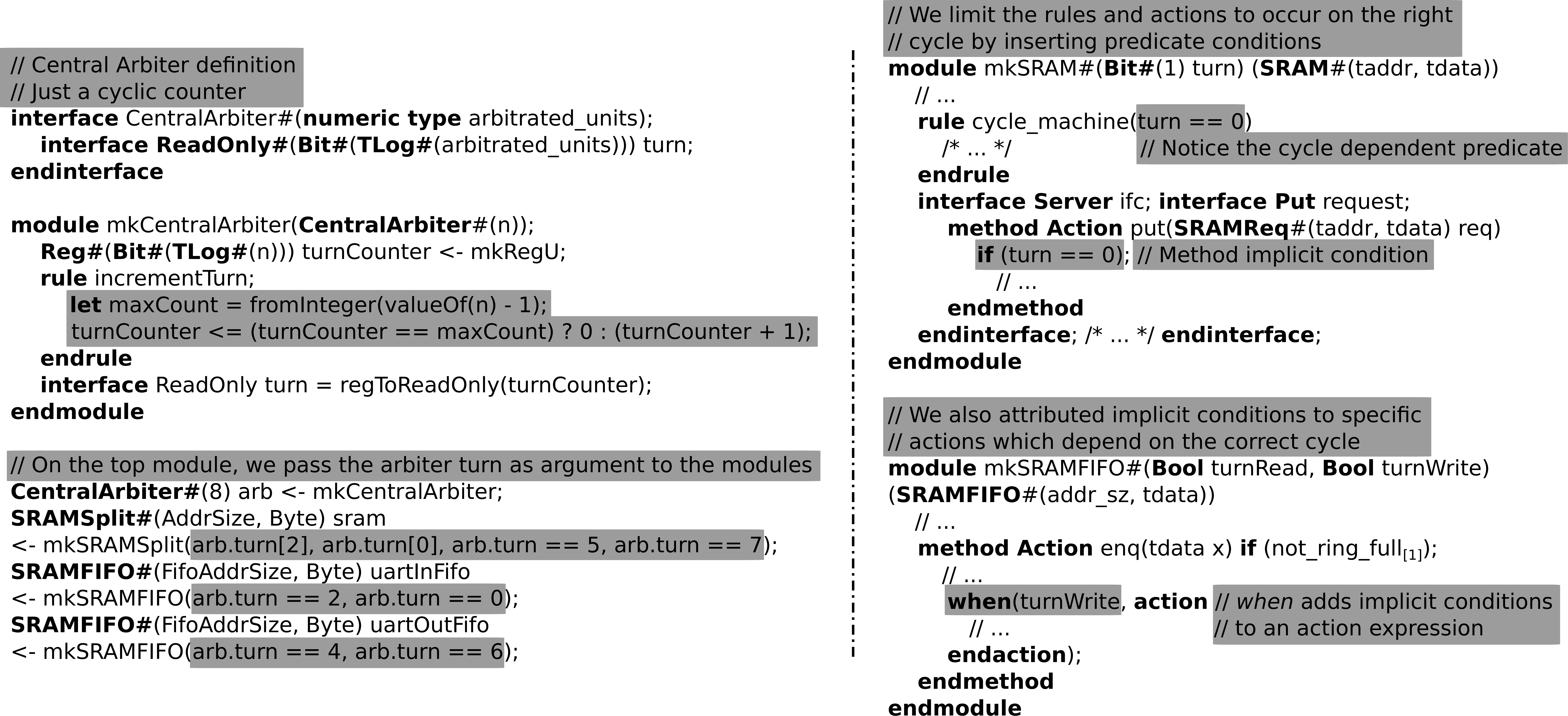}}
\caption{Excerpts from the static arbiter implementation. The \texttt{CentralArbiter} module, which consists of a simple cyclic counter, is instantiated inside the top-level module. Its \texttt{turn} method describes the current cycle according to the schedule of Figure~\ref{DiagramaTempo}. Some bits of the cycle counter, or Boolean conditions involving its current state, are routed to inner modules, where they are appended to predicates or added as implicit conditions.}
\label{StaticArbiter}
\end{center}
\end{figure*}

We added these predicate conditions based on a manually devised arbitration schedule, shown in Figure~\ref{DiagramaTempo}. This schedule allows for the execution of an enqueue and a dequeue operation on both \texttt{uartInFifo} and \texttt{uartOutFifo} during the course of 8~clock cycles. A central arbiter, which consists of a counter reset every 8~cycles, was implemented just below the top level module. Boolean values derived from this counter, signaling if each operation could occur during each cycle, were routed from the top level module to the inner SRAM controller, \texttt{SRAMSplit} and \texttt{SRAMFIFO}s (Figure~\ref{StaticArbiter}). After the predicates were changed, some FIFOs could be removed, reducing the number of LEs needed to implement the design.

\textbf{SRAM controller}: The \texttt{cycle} register (compare with Figure~\ref{BlocosSRAM2CLKDyn}) was removed and replaced by the least significant bit of the central arbiter counter. Memory requests became allowed only when this bit is zero, which happens in cycles numbered 0, 2, 4 and 6 (Figure~\ref{DiagramaTempo}).

\textbf{SRAMSplit}: Requests to one of the arbitrated servers became allowed only during the correct cycle, as defined by the timing diagram --- requests coming from \texttt{uartInFifo} at cycles numbered 0 and 2, and those from \texttt{uartOutFifo} at cycles 4 and 6. Both \texttt{reqfifo}s could then be removed from the design without affecting its behavior. The order of responses could also be inferred from the diagram (cycles 3, 5, 7 and 1), allowing us to remove the \texttt{pending} FIFO. After all changes, the module became just an abstraction which synthesizes purely to wires (compare with Figure~\ref{BlocosSRAMSplitDyn}).

\textbf{SRAMFIFO}: Dequeues and enqueues became allowed only during the designated cycles (2 and 4 for dequeues, 0 and 6 for enqueues). This allowed to remove memory request output FIFOs, which were replaced by wires.

\section{Results}\label{Results}

\subsection{Synthesis}

Synthesis results are shown in Table~\ref{SynthesisResults}. On the device actually adopted in our project (EPM\-2210\-F324\-C3), the dynamic arbitrated circuit occupies 200 more LEs than the static arbiter design. This corresponds to 9\% of the LEs available in the CPLD. Almost a half of the hardware resources are still free and could be exploited to implement new features. We have also synthesized both architectures on a smaller device (EPM\-1270\-F256\-C3) in order to demonstrate the design can meet the requirements even when reaching the limits of the CPLD substrate. The Quartus II Fitter clearly undertook more effort during synthesis on this device, as both circuits were implemented occupying less LEs than on the bigger CPLD. Nonetheless, the attainable clock frequency was not significantly reduced by this area optimization, remaining above 50~MHz.

\begin{table}[htbp]
\centering
\caption{Synthesis results for both arbiter designs (Altera Quartus II 14.0)}
\label{SynthesisResults}
\begin{tabular}{|c||c|c||c|c|} \hline
& \multicolumn{2}{|c||}{EPM2210F324C3} & \multicolumn{2}{|c|}{EPM1270F256C3} \\ \hline \hline
\textbf{Design} & \textbf{Logic}    & \textbf{Maximum clock} & \textbf{Logic}    & \textbf{Maximum clock} \\
\textbf{Arbiter} & \textbf{elements} & \textbf{frequency}     & \textbf{elements} & \textbf{frequency} \\ \hline
Static & 1017 (46\%) & 54.57~MHz & 970 (76\%) & 54.36~MHz \\ \hline
Dynamic & 1217 (55\%) & 54.07~MHz & 1168 (92\%) & 53.49~MHz \\ \hline
\end{tabular}
\end{table}

\subsection{Experimental validation}

\begin{figure*}[tbp]
\centering
\includegraphics[width=4.6in]{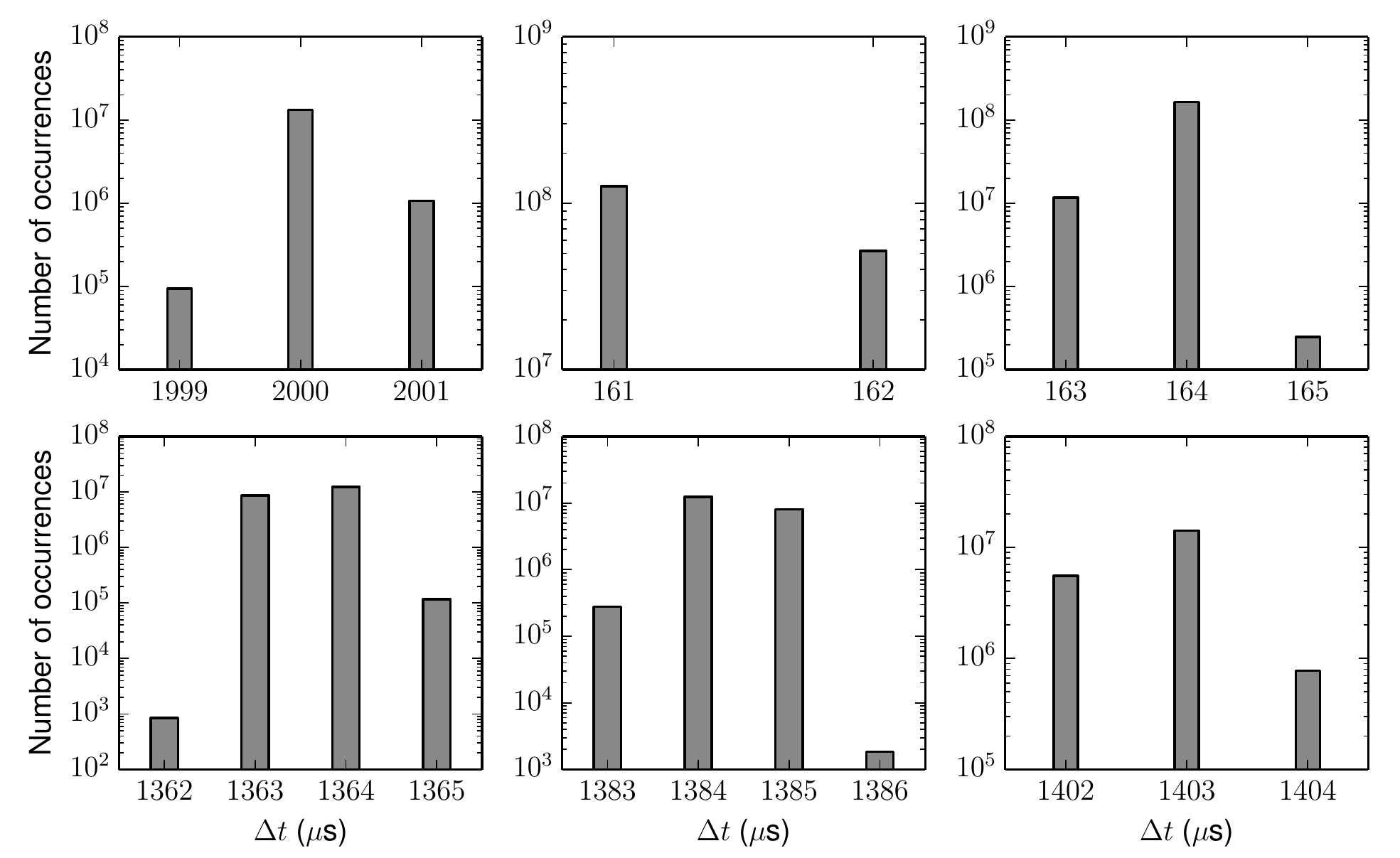}
\caption{Histograms displaying the number of occurrences of each time interval $\Delta t$ measured between two consecutive pulses. Each one of the six inputs channels was connected to an independent periodic signal source. No missed nor spurious events could be observed even after eight hours of acquisition.}
\label{histograms_8h_acq}
\end{figure*}

Workbench validation consisted in connecting independent square-wave periodic signal generators into each input of the system for 8~hours and then analyzing the acquired data to look for spurious or missing detections. Figure~\ref{histograms_8h_acq} shows histograms of the time interval ($\Delta t$) between two consecutive recorded pulses for all input channels. Histograms shown in the first line correspond to periodic pulses generated by an 1-chan\-nel Hewlett-Packard 33120A and a 2-chan\-nel Sony-Tektronix AFG320 function generator. The three remaining channels were fed with signals generated by free-running astable oscillators made using the NE555 timer integrated circuit.

The first input channel was programmed to synchronize DAC conversions, thus we have fed it with 500~Hz (frame rate frequency adopted by the VSImG \cite{deAlmeida20111762} visual stimulation system). The second and third channels were supplied with close but incommensurable frequencies (6.2~kHz and 6.1~kHz). The remaining channels were fed by similar frequencies, produced by three identical NE555 circuits, differing only within component nominal tolerances.

Experiments with both architectures (dynamic and static arbiter) resulted in almost identical histograms, thus the figure only portrays the results for one of them (dynamic arbiter). The histograms show that during 8~hours of acquisition no pulses were missed, and no spurious events were registered, otherwise the abscissa of the graph would reach double or half the value of the baseline period, respectively. The maximum deviation from the adjusted periods was within the acceptable generator's thermal drift. As expected, NE555 oscillators are less stable and produce more jitter than the commercial function generators, resulting in sparser histograms.

We emphasize that input event periods employed during this test were well below the minimum intervals between spikes (refractory period) attainable by a typical neuron. For example, in \textit{Chrysomya megacephala}'s H1 neuron this minimum interval is 2~ms \cite{Baptista13022008}.

\subsection{Arbiter resilience evaluation}

\begin{figure*}[htbp]
\centering
\includegraphics[width=4.6in]{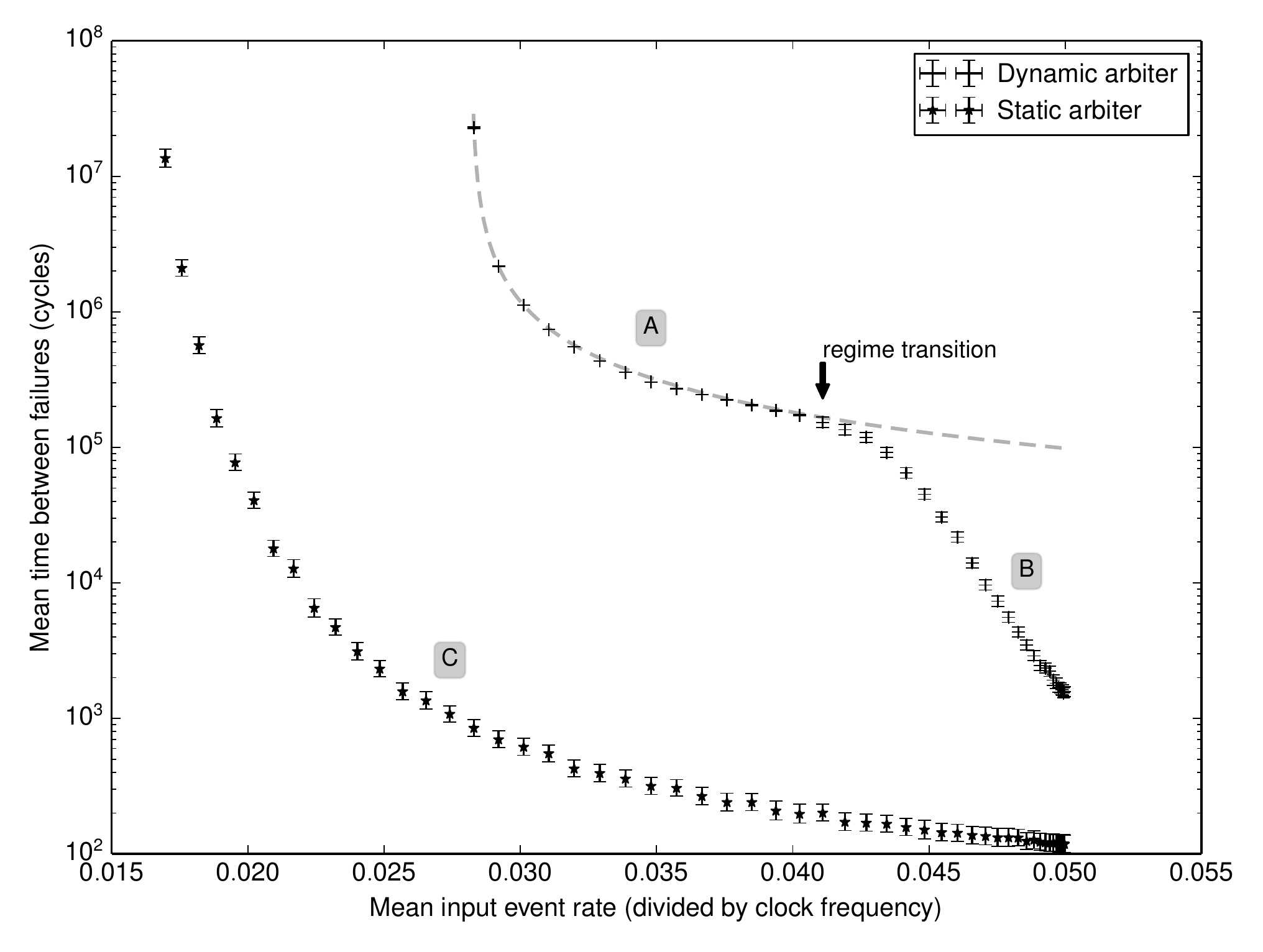}
\caption{Mean time between failures (MTBF) obtained by simulation, with designs configured for an increased timestamp resolution of~$\nicefrac{1}{20}$ of the system clock, and subject to large mean input event rates. The dynamic arbiter is consistently more resilient, and presented two operating regimes: in regime \textbf{A}, failures are caused by overflow of \texttt{uartOutFifo}, while in regime \textbf{B}, memory write requests occur at a high rate and eventually cannot be scheduled on time, overflowing \texttt{funnel}. As the static arbiter schedules enqueuing operations at a fixed rate of 1~byte per 8~cycles (slower than the simulated UART transmission rate), regime \textbf{C} is only due to \texttt{funnel} overflow.}
\label{mtbf}
\end{figure*}

Besides the dynamic arbiter design advantages related to code reusability, inherent to its latency insensitive and decoupled characteristics, it is also more resilient to failures. In order to prove this, we needed to increase the timestamp resolution above its original specification of 1~$\mu$s. In fact, any update rate below~$\nicefrac{1}{40}$ of the clock frequency (50~MHz) can always be correctly scheduled, as it leaves room for at least 5~rounds of 4~memory operations, each one taking 2~cycles (see Figure~\ref{DiagramaTempo}), sufficient to carry the 5-byte (\texttt{channelFlags}, \texttt{timestamp}) tuple in and out of the FIFOs. Thus, to be able to observe failures related to differences in SRAM arbitration schemes, we increased the timestamp counter update rate from~$\nicefrac{1}{50}$ to~$\nicefrac{1}{20}$ of the clock frequency.

In order to keep control over parameters such as UART transmission rate, we simulated the system instead of evaluating it with workbench instruments. Both architectures were simulated under exactly the same parameters and inputs. Inputs were fed with trains of pulses generated according to a Poisson process, a stochastic model that occasionally produces activity bursts, although it possesses a parameterized mean rate. It has also been adopted in some statistical models of spiking neurons \cite{Dayan:2005:TNC:1205781}. The first channel, however, was modeled as an oscillation whose frequency varies according to a narrow Gaussian distribution, which better reproduces the behavior of the stimuli reference clock.

We limited UART transmission to 1~byte per 6~cycles, aiming to observe the system in a regime in which it would eventually acquire more data than it could transmit. UART reception was constrained to 1~byte per 10~cycles, allowing 3-byte chunks of stimuli data to be provided to the system at double the speed required by the first-channel reference clock, whose mean frequency was chosen at~$\nicefrac{1}{60}$ of the system clock.

Varying the rate parameter of the Poisson processes, we measured the total mean input event rate to which the system was exposed, i.e. the number of input events divided by the number of cycles of simulation, relating it to the mean time between failures (MTBF) in number of cycles. Failures were detected if any of the four error conditions discussed in Section \ref{BSVmod} were triggered. In this experiment they happened only due to overflows in the \texttt{tx} path.

Figure~\ref{mtbf} shows these results and demonstrates that besides supporting a mean firing rate greater than the static arbiter without missing events, the dynamic arbiter takes longer to fail when the frequency approaches its limits. At an input rate of $\nicefrac{1}{20}$, the maximum meaningful frequency at the adopted timestamp resolution, the dynamic arbiter fails after circa $10^3$ cycles, whereas the static arbiter withstands for only $10^2$~cycles.

Under the simulation parameters, the static arbiter is not able to fill \texttt{uart\-Out\-Fifo}'s SRAM-contained ring with more than 1~byte. This happens because the FIFO enqueuing rate is limited to a maximum of 1~byte per 8~cycles by the central arbiter schedule (see Figure~\ref{DiagramaTempo}), while the simulated UART can reach a transmission rate of 1~byte per 6~cycles, sufficient to keep the FIFO almost empty. The observed failures were due to overflow of the \texttt{funnel}: the \texttt{uartOutFifo} never even came close to a full state. Therefore, the failure process may be viewed as nearly stationary at a time scale greater than $10^2$~cycles. Indeed, the measured data set (Figure~\ref{mtbf}--C) does not change significantly if we reset the circuit state a couple of times during the course of simulation.

On the other hand, the dynamic arbiter is able to surpass an operating regime (Figure~\ref{mtbf}--B) where failures occur because of \texttt{funnel} overflow, reaching another regime at lower frequencies (Figure~\ref{mtbf}--A) where the system does not abort until \texttt{uartOutFifo} is full. The dashed curve is proportional to $(f - f_\mathrm{lim})^{-1}$, where $f$ is the mean input frequency (abscissa) and $f_\mathrm{lim}$ is a limit frequency (in this experiment, $f_\mathrm{lim} \approx 0.0282$) such that $5 f_\mathrm{lim}$ is less than the effective UART transmission rate, implying the circuit is not expected to ever fail for $f \le f_\mathrm{lim}$.

\section{Conclusions}\label{Conclusion}

The system described in this paper (source code at \cite{Matias:11034}) can be applied to neuroscience research both on \textit{in vivo} or \textit{in vitro} experiments requiring deterministic timing and synchronous stimuli generation, such as the study of neural coding on the visual system of flies \cite{6796624}. It can also be applied to experiments in neuroethology, for example on the analysis of electrocommunication signals produced by pulse-type electric fish \cite{10.1371/journal.pone.0084885,Matias2014,ForlimdeAlmeida2012,Guariento2014}. The employed digital pulse timestamping technique allows to achieve a measurement precision in the order of 1~$\mu$s, much higher than most ADC-based acquisition systems. Even though our project was programmed to a small reconfigurable device, almost a half of LEs were left free and can be filled to implement future experiments with real-time feedback \cite{Muniz2009}.

We have also shown that Bluespec SystemVerilog (BSV) can be effectively used even in projects involving small devices, and have presented an approach to refactor a decoupled and latency insensitive logic into a statically arbitrated one, which could be useful when a designer needs to quickly lower a system's LE usage --- however, keeping dynamic arbitration can compensate the LE cost if the system needs to be resilient to activity bursts. The designed code is modular and reusable to implement similar systems, e.g. we have a working prototype for closed-loop experiments implemented on an EP4\-CGX\-150\-DF31\-C7 FPGA, which occupies 5728~LEs (4\% of the device) and interfaces with a real-time operating system (RTOS) through PCI Express \cite{2015arXiv150400932O}.


\subsubsection*{Acknowledgments.}{\small Authors were supported by grants from CAPES and FAPESP. Maxim Integrated provided free analog IC samples. Altera Corp and Bluespec Inc supplied free software licenses through their university programs.}

\bibliographystyle{splncs}
\bibliography{referencias}
\end{document}